\documentclass[twocolumn,showpacs,aps]{revtex4}
\usepackage{graphicx}
\usepackage{dcolumn}
\usepackage{amsmath}
\usepackage{epsfig}
\usepackage{axodraw}
\usepackage{color}

\begin{document}

\title{\large \bfseries \boldmath Radiative and leptonic decays of the pseudoscalar charmonium state $\eta_c$}
\author{Mao-Zhi Yang}\email{yangmz@nankai.edu.cn}
\affiliation{Department of Physics, Nankai University, Tianjin,
300071, China \footnote{Mailing address}}
\affiliation{Kavli
Institute for Theoretical Physics China, CAS, Beijing, 100190,
China}




\begin{abstract}
The radiative and leptonic decays of $\eta_c\to \gamma\gamma$ and
$\eta_c\to l^+l^-$ are studied. For $\eta_c\to \gamma\gamma$
decay, the second-order electromagnetic tree-level diagram gives
the leading contribution. The decay rate of $\eta_c\to
\gamma\gamma$ is calculated, the prediction is in good agreement
with the experimental data. For $\eta_c\to l^+\l^-$, both the tree
and loop diagrams are calculated. The analysis shows that the loop
contribution dominates, the contribution of tree diagram with
$Z^0$ intermediate state can only modifies the decay rate by less
than 1\%. The prediction of the branching ratios of $\eta_c\to
e^+e^-$ and $\mu^+\mu^-$ are very tiny within the standard model.
The smallness of these predictions within the standard model makes
the leptonic decays of $\eta_c$ sensitive to physics beyond the
standard model. Measurement of the leptonic decay may give
information of new physics.
\end{abstract}


\maketitle

The radiative and leptonic decays of $\eta_c\to \gamma\gamma$ and
$\eta_c\to l^+l^-$ involve electromagnetic and weak interactions.
The decay rates are determined not only by electroweak
interaction, but also by strong interaction, which binds the
quark-antiquark pair $c\bar{c}$ in $\eta_c$ together. The decay
amplitudes of these decay processes are generally convolutions of
the wave function of $\eta_c$ and the electroweak transition
amplitude of $c\bar{c}\to \gamma\gamma$ and/or $c\bar{c}\to
l^+l^-$. The decay process of $\eta_c\to\gamma\gamma$ can be used
to test the decay constant and the wave function of $c\bar{c}$ in
this meson state. The decay rate of $\eta_c\to\gamma\gamma$ has
been measured in experiment \cite{PDG2008}. Within the framework
of the standard model, the leptonic decay $\eta_c\to l^+l^-$ is
contributed by fourth-order electromagnetic transition and
tree-level weak transition induced by $Z^0$. The low probability
of a fourth-order electromagnetic and weak transition makes the
leptonic decays sensitive to hypothetical interactions arising
from physics beyond the standard model, such as the existence of a
light pseudoscalar Higgs boson in the next-to-minimal
supersymmetric standard model \cite{higgs,gh,fs,htv,cy} or
leptoquark bosons that carry both quark and lepton flavors
\cite{gg,ps,eh,bw}, both of which can enhance the leptonic decays
of $\eta_c$ with appropriate values of new physics parameters in
these models.

In the literature there have been many theoretical works on the
calculations of two-photon decays of pseudoscalar heavy quarkonia
$\eta_c$, $\eta_b$ etc., based on relativistic quark model or
potential model \cite{am,efg,hw}, Bethe-Salpeter equation
\cite{chlt, klw}, heavy-quark spin symmetry \cite{lp}, and Lattice
QCD \cite{de}. However the decay rates of leptonic decays of
pseudoscalar heavy quarkonia have not been known yet.

In this work the radiative and leptonic decays of $\eta_c\to
\gamma\gamma$ and $\eta_c\to l^+l^-$ are studied consistently
within the framework of the standard model, by using a method
different from those used in the literature.  The effective
Hamiltonians in quark level for these decays are calculated at
first. Then the factorization formula is derived, the decay
amplitudes are expressed as convolutions of meson wave function
and the hard transition amplitudes, where the wave function is
controlled by nonperturbative QCD, for which I use the result
calculated in QCD sum rule \cite{wv}, while the hard transition
amplitudes are calculated with perturbation theory. For the
process $\eta_c\to l^+l^-$, the loop diagrams are calculated
analytically. The infrared divergence is analyzed, the possible
information of new physics is also briefly discussed.

\begin{figure}[hbt]
\begin{center}
\vspace{0.5CM}
 \epsfig{file=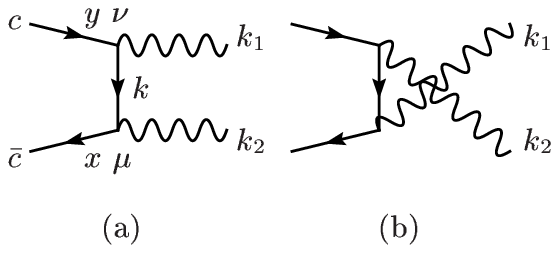, width=6cm,height=2.5cm}
 \end{center}
\vspace{-0.5cm} \caption{\small Diagrams for $\eta_c\to
\gamma\gamma$. } \label{fig1}
\end{figure}

The diagram for the decay of $\eta_c\to \gamma\gamma$ is depicted
in Fig.\ref{fig1}. The transition matrix element relevant to
Fig.\ref{fig1}(a) can be written in coordinate space as
\begin{eqnarray}\label{eq1}
&&T_1=\langle \gamma\gamma |\int d^4xd^4y \bar{c}(x)iQ_c
e\gamma_\mu\int\frac{d^4k}{(2\pi)^4}\\
\nonumber &&\cdot \frac{i}{\not{k}-m_c}e^{-ik\cdot
(x-y)}iQ_ce\gamma_\nu c(y)A^\mu(x)A^\nu (y)|\eta_c\rangle ,
\end{eqnarray}
where $e$ is the absolute value of the charge of electron, $Q_c$
the charge of the $c$ quark in unit of $e$, $A^\mu(x)$ and $A^\nu
(y)$ are the electromagnetic fields for the two photons.
Contracting the creation operators of the two photons with that in
the electromagnetic fields, then the amplitude $T_1$ becomes
\begin{eqnarray}\label{eq2}
&&T_1=\int d^4xd^4y\int\frac{d^4k}{(2\pi)^4}e^{-ik\cdot (x-y)}\\
\nonumber &&\cdot  [iQ_c e\gamma_\mu
\frac{i}{\not{k}-m_c}iQ_ce\gamma_\nu]_{\alpha\beta}\epsilon^{*\mu}_1
e^{ik_1x} \epsilon^{*\nu}_2 e^{ik_2y}\\ \nonumber &&\cdot \langle
0 | \bar{c}(x)_\alpha c(y)_\beta |\eta_c\rangle,
\end{eqnarray}
where $\alpha$ and $\beta$ are the Dirac spinor indices. It is
understood that the repeated indices are summed. In both eqs.
(\ref{eq1}) and (\ref{eq2}), summations over color degrees of
freedom and a factor $P\mbox{exp}\{i\int dz^\mu T^a A^a_\mu (z)\}$
between $\bar{c}(x)$ and $c(y)$ that makes the amplitude
gauge-invariant are indicated.

The leading-twist wave function for $\eta_c$ meson can be defined
through the matrix element $\langle 0| \bar{c}(x)_\alpha
c(y)_\beta |\eta_c\rangle$ \cite{cz,bf}, which will be further
discussed in Appendix A,
\begin{eqnarray} \label{wv}
&&\langle 0| \bar{c}(x)_\alpha c(y)_\beta |\eta_c\rangle
\\\nonumber &&=-\frac{i}{4}f_{\eta_c}\int_0^1 du e^{-i(up\cdot x+\bar{u}p\cdot
y)}[\not {p}\gamma_5]_{\beta\alpha}\phi(u,\mu),
\end{eqnarray}
where $\bar{u}=1-u$, $\mu$ is an energy scale, and $f_{\eta_c}$ is
the decay constant of $\eta_c$, which is defined as
\begin{equation}
\langle 0| \bar{c}\gamma_\mu\gamma_5 c|\eta_c\rangle
=if_{\eta_c}p_\mu,
\end{equation}
here $p_\mu$ is the four-momentum of the meson.

With the matrix element given in eq.(\ref{wv}), the transition
matrix element $T_1$ becomes
\begin{eqnarray}\label{eq5}
&&T_1=-\frac{1}{4}f_{\eta_c}Q_c^2e^2\int_0^1du\int
d^4xd^4y\int\frac{d^4k}{(2\pi)^4}
\\
\nonumber &&\cdot e^{-ik\cdot (x-y)}e^{ik_1\cdot x}e^{ik_2\cdot
y}e^{-i(up\cdot x+\bar{u}p\cdot y)}\phi(u,\mu)\epsilon^{*\mu}_1
 \epsilon^{*\nu}_2\\
\nonumber &&\frac{1}{k^2-m_c^2}Tr[\not{p}\gamma_5\gamma_\mu
(\not{k}+m_c)\gamma_\nu].
\end{eqnarray}
It is not difficult to perform the integration over the
coordinates $x$, $y$ and the momentum $k$. After these
manipulations, one can obtain
\begin{eqnarray}
&&T_1=if_{\eta_c}Q^2_ce^2\int_0^1du
\frac{\phi(u,\mu)}{(\bar{u}k_1-uk_2)^2-m_c^2}\\\nonumber &&
\cdot\epsilon_{\rho\mu\sigma\nu}
k_1^{\rho}k_2^{\sigma}\epsilon^{*\mu}_1\epsilon^{*\nu}_2
(2\pi)^4\delta^4(k_1+k_2-p).
\end{eqnarray}
Remove the overall four-momentum conservation factor
$(2\pi)^4\delta^4(k_1+k_2-p)$, one can obtain the decay amplitude
contributed by Fig.\ref{fig1}(a)
\begin{eqnarray}
&&A_1=if_{\eta_c}Q^2_ce^2\int_0^1du
\frac{\phi(u,\mu)}{(\bar{u}k_1-uk_2)^2-m_c^2}\\\nonumber &&
\cdot\epsilon_{\rho\mu\sigma\nu}
k_1^{\rho}k_2^{\sigma}\epsilon^{*\mu}_1\epsilon^{*\nu}_2.
\end{eqnarray}
The contribution of Fig.\ref{fig1}(b) can be obtained by making
the exchange of $k_1\leftrightarrow k_2$ and
$\epsilon_1\leftrightarrow \epsilon_2$ in $A_1$
\begin{equation}
A_2=A_1{\left(\begin{array}{c}
                 \epsilon_1\leftrightarrow \epsilon_2\\
                 k_1\leftrightarrow k_2
                \end{array} \right)}.
\end{equation}
Then the total contribution of Fig.\ref{fig1} (a) and (b) is
\begin{equation}
A=A_1+A_2.
\end{equation}
Using $k_1^2=k_2^2=0$ and $2k_1\cdot k_2=m^2_{\eta_c}$, one can
simplify the amplitude $A$ to be
\begin{eqnarray}
&&A=2if_{\eta_c}Q^2_ce^2\int_0^1du
\frac{\phi(u,\mu)}{u\bar{u}m^2_{\eta_c}+m_c^2}\\\nonumber &&
\cdot\epsilon_{\rho\sigma\mu\nu}
k_1^{\rho}k_2^{\sigma}\epsilon^{*\mu}_1\epsilon^{*\nu}_2.
\end{eqnarray}
For simplicity in the following calculation, one can define a new
quantity $M$ as
\begin{equation}
M\equiv 2if_{\eta_c}Q^2_ce^2\int_0^1du
\frac{\phi(u,\mu)}{u\bar{u}m^2_{\eta_c}+m_c^2},
\end{equation}
then we can obtain the square of the total amplitude
\begin{equation}
|A|^2=\frac{1}{2}m^4_{\eta_c}|M|^2.
\end{equation}
The decay width can be calculated by the following formula
\begin{equation}
\Gamma(\eta_c\to \gamma\gamma)=\frac{1}{2!}\frac{1}{8\pi}|A|^2
\frac{|\vec{k}|}{m^2_{\eta_c}},
\end{equation}
where $\frac{1}{2!}$ is the statistic factor for two identical
particles, $\vec{k}$ is the three-momentum of one of the photons
in the rest frame of $\eta_c$.

\begin{figure}[hbt]
\begin{center}
\vspace{1CM}
 \epsfig{file=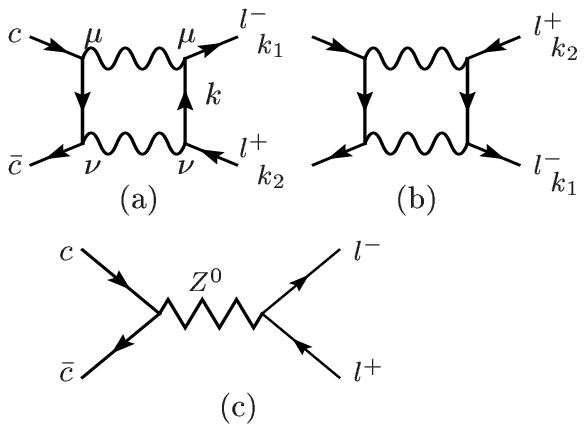, width=6cm,height=4.5cm}
 \end{center}
\vspace{-0.5cm} \caption{\small Diagrams for $\eta_c\to l^+l^-$.
(a) and (b) QED contribution where two virtual photons as
intermediate state, (c) weak $Z^0$ contribution.} \label{fig2}
\end{figure}

Next let us discuss the leptonic decay $\eta_c\to l^+l^-$. In the
standard model, the Feynman diagrams for this process include
photon-Fermion loop diagrams and $Z^0$ tree diagram, which are
depicted in Fig.\ref{fig2}.

The effective Hamiltonian describing the transition of
$c\bar{c}\to l^+l^-$ can be calculated with the help of the
Feynman diagram in Fig.\ref{fig2}. The contribution of
Fig.\ref{fig2} (a) is calculated to be
\begin{eqnarray}
&&H_1=Q_c^2e^4\int\frac{d^4k}{(2\pi)^4}
\\ \nonumber
&&\cdot\frac{\bar{l}\gamma^\mu
 (\not{k}+m_l)\gamma^\nu l~\bar{c}\gamma_\nu
 (\not{k}-\not{k}_1+\not{p}_c+m_c)\gamma_\mu
 c}{\{
 (k^2-m_l^2+i\epsilon)[(k-k_1)^2+i\epsilon]
 [(k+k_2)^2+i\epsilon]}\\\nonumber
 &&\hspace{3cm }\times[(k-k_1+p_c)^2-m_c^2+i\epsilon]\},
\end{eqnarray}
where $m_l$ is the lepton mass, $p_c$ the four-momentum of $c$
quark. The quark fields $c$ and $\bar{c}$ are in the momentum
space. They are related to the field operators in the coordinate
space by
\begin{eqnarray}
&&\bar{c}=\int d^4x
\bar{c}(x)e^{ip_{\bar{c}}\cdot x},\\
&&c=\int d^4y c(y)e^{ip_c\cdot y}.
\end{eqnarray}
Then the effective Hamiltonian $H_1$ can be reexpressed in terms
of quark fields in the coordinate space
\begin{eqnarray}
&&H_1=\int d^4xe^{ip_{\bar{c}}\cdot x}\int d^4ye^{ip_c\cdot y}
Q_c^2e^4\int\frac{d^4k}{(2\pi)^4}
\\ \nonumber
&&\times\frac{\bar{l}\gamma^\mu
 (\not{k}+m_l)\gamma^\nu l~\bar{c}(x)\gamma_\nu
 (\not{k}-\not{k}_1+\not{p}_c+m_c)\gamma_\mu
 c(y)}{a_k},
\end{eqnarray}
where $a_k$ is defined as
\begin{eqnarray}
&&a_k \equiv (k^2-m_l^2+i\epsilon)[(k-k_1)^2+i\epsilon]
 \\\nonumber
&& ~~~\times[(k+k_2)^2+i\epsilon][(k-k_1+p_c)^2-m_c^2+i\epsilon].
\end{eqnarray}
With the effective Hamiltonian $H_1$, the amplitude contributed by
Fig.\ref{fig2} (a) is
\begin{eqnarray}\label{a1}
A_1&=&\int\frac{d^4p_c}{(2\pi)^4}\frac{d^4p_{\bar{c}}}{(2\pi)^4}
\langle l^+l^-|H_1|\eta_c\rangle
\\\nonumber &=&Q^2_c
e^4 \int\frac{d^4p_c}{(2\pi)^4}\frac{d^4p_{\bar{c}}}{(2\pi)^4}
\int d^4xe^{ip_{\bar{c}}\cdot x}\int d^4ye^{ip_c\cdot
y}\\\nonumber &&
\cdot\int\frac{d^4k}{(2\pi)^4}\bar{u}(k_1)\gamma^\mu
(\not{k}+m_l)\gamma^\nu v(k_2)\\\nonumber &&\times\langle
0|\bar{c}(x)\gamma_\nu
 (\not{k}-\not{k}_1+\not{p}_c+m_c)\gamma_\mu c(y)|\eta_c\rangle
 \frac{1}{a_k},
\end{eqnarray}
The matrix element in the above equation can be treated as
\begin{eqnarray}\label{e20}
&& \langle 0|\bar{c}(x)\gamma_\nu
 (\not{k}-\not{k}_1+\not{p}_c+m_c)\gamma_\mu c(y)|\eta_c\rangle
 \\\nonumber
 &&=\langle 0|\bar{c}(x)_\alpha c(y)_\beta|\eta_c\rangle [\gamma_\nu
 (\not{k}-\not{k}_1+\not{p}_c+m_c)\gamma_\mu]_{\alpha\beta}.
\end{eqnarray}
With the help of eq.(\ref{e20}) and the wave function of $\eta_c$
defined in eq.(\ref{wv}), the decay amplitude $A_1$ can be
expressed as
\begin{eqnarray}
&&A_1=Q^2_c e^4\int\frac{d^4k}{(2\pi)^4}\bar{u}(k_1)\gamma^\mu
(\not{k}+m_l)\gamma^\nu v(k_2)\\\nonumber
&&~~\times\frac{-i}{4}f_{\eta_c}\int_0^1\phi(u,\mu)
\mbox{Tr}[\not{p}\gamma_5\gamma_\nu
(\not{k}-\not{k}_1+\bar{u}\not{p}\\\nonumber
&&~~~~+m_c)\gamma_\mu]\frac{1}{a_k}.
\end{eqnarray}
Perform the trace operation and using the identity of the gamma
matrices
\begin{equation}
\gamma_\alpha\gamma_\beta\gamma_\lambda
=g_{\alpha\beta}\gamma_{\lambda}+g_{\beta\lambda}\gamma_{\alpha}
-g_{\alpha\lambda}\gamma_{\beta}
+i\epsilon_{\mu\alpha\beta\lambda}\gamma^\mu\gamma_5,
\end{equation}
the amplitude $A_1$ can be further calculated to be
\begin{eqnarray}\label{aa1}
&&A_1=iQ_c^2e^4f_{\eta_c}\int_0^1du\phi(u,\mu)
[\bar{u}(k_1)\gamma^\alpha\gamma_5\\\nonumber &&~~\cdot
v(k_2)a_{1\alpha}
-m_l\bar{u}(k_1)\sigma^{\mu\nu}v(k_2)\epsilon_{\mu\nu\rho\sigma}p^\rho
a_2^\sigma],
\end{eqnarray}
where
\begin{eqnarray}\label{aacoe}
&&a_{1\alpha}=2\int\frac{d^4k}{(2\pi)^4}\\\nonumber
 &&~~~~~\times\frac{ p_\alpha
k\cdot (k-k_1+\bar{u}p)-(k-k_1+\bar{u}p)_\alpha p\cdot k}{a_k},\\
\label{aacoe2}
&&a_2^\sigma=\int\frac{d^4k}{(2\pi)^4}\frac{(k-k_1+\bar{u}p)^\sigma}{a_k}.
\end{eqnarray}
Next one needs to perform the loop integrations in the
coefficients $a_{1\alpha}$ and $a_2^\sigma$.

Using $p=k_1+k_2$ and a few steps of algebra manipulation, one can
cancel one propagator in the coefficient $a_{1\alpha}$, then the
four-point loop integration of $a_{1\alpha}$ can be reduced to a
sum of a few three-point loop integrals
\begin{widetext}
\begin{eqnarray}
&&a_{1\alpha}=\int\frac{d^4k}{(2\pi)^4}\left\{\frac{}{}\right.
p_\alpha\times \left[\frac{1}{[(k-k_1)^2+i\epsilon]
 [(k+k_2)^2+i\epsilon][(k-k_1+\bar{u}p)^2-m_c^2+i\epsilon]}\right.
 \\\nonumber &&
 +\frac{u}{(k^2-m_l^2+i\epsilon)[(k+k_2)^2+i\epsilon]
 [(k-k_1+\bar{u}p)^2-m_c^2+i\epsilon] }
 +\left.\frac{\bar{u}}{(k^2-m_l^2+i\epsilon)[(k-k_1)^2+i\epsilon]
 [(k-k_1+\bar{u}p)^2-m_c^2+i\epsilon]}\right]
 \\\nonumber  &&
 -\frac{(k-uk_1+\bar{u}k_2)_\alpha}{(k^2-m_l^2+i\epsilon)[(k-k_1)^2+i\epsilon]
 [(k-k_1+\bar{u}p)^2-m_c^2+i\epsilon]}
 +\left.\frac{(k-uk_1+\bar{u}k_2)_\alpha}{(k^2-m_l^2+i\epsilon)
 [(k+k_2)^2+i\epsilon]
 [(k-k_1+\bar{u}p)^2-m_c^2+i\epsilon] }
 \right\}.
\end{eqnarray}
\end{widetext}
With the Feynman parameterization, the integration
over the loop momentum in the above equation can be performed,
then the coefficient $a_{1\alpha}$ can be expressed as integrals
over Feynman parameters
\begin{eqnarray}\label{a1a}
a_{1\alpha}&=&\frac{-i}{16\pi^2}\int_0^1dx\int_0^{1-x}dy
\left\{p_\alpha\left[\frac{1}{a_1}+\frac{u}{a_2}\right.\right.\\\nonumber
&&+\left.\frac{\bar{u}}{a_3}\right]
-\frac{[y+u(x-1)]k_{1\alpha}+\bar{u}\bar{x}k_{2\alpha}}{a_3}\\\nonumber
&&
+\left.\frac{-u\bar{x}k_{1\alpha}+(\bar{u}-\bar{u}x-y)k_{2\alpha}}{a_2}\right\},
\end{eqnarray}
with
\begin{eqnarray}\label{a1to3}
&&
a_1=m_{\eta_c}^2[\bar{u}x+y)(\bar{u}x+y-1)+xu\bar{u}]+xm_c^2\\\nonumber
&&~~~~~-i\epsilon,\\\nonumber &&
a_2=[1-2x-2y+(x+y)^2]m_l^2+x
u(\bar{x}\bar{u}-
   y)m_{\eta_c}^2\\\nonumber &&~~~~~-i\epsilon,\\\nonumber &&
a_3=[1-2x-2y+(x+y)^2]m_l^2+x\bar{u}(\bar{x}u-
   y)m_{\eta_c}^2\\\nonumber &&~~~~~-i\epsilon.
\end{eqnarray}
While the four-point loop integration of the coefficient
$a_2^\sigma$ in eq.(\ref{aacoe2}) can not be reduced, it can be
expressed as a 3-fold integral of Feynman parameters
\begin{eqnarray}\label{a2s}
a_2^\sigma &=& \frac{i}{16\pi^2}\int_0^1dx\int_0^{1-x}dy
\int_0^{1-x-y}dz\\\nonumber
&&\times\frac{(ux+z-u)k_1^\sigma-(\bar{u}x+y-\bar{u})k_2^\sigma}{a_4^2},
\end{eqnarray}
where
\begin{eqnarray}\label{a4}
a_4&=&(1-x-y-z)^2m_l^2+[xu\bar{u}\\\nonumber
  &&-(ux+z)(\bar{u}x+y)]m_{\eta_c}^2+xm_c^2-i\epsilon .
\end{eqnarray}
Substitute $a_{1\alpha}$ and $a_2^\sigma $ in eqs.(\ref{a1a}) and
(\ref{a2s}) into eq.(\ref{aa1}), the amplitude $A_1$ can be
reduced to
\begin{eqnarray}\label{ax1}
&&A_1=\frac{Q_c^2e^4}{16\pi^2}f_{\eta_c}\int_0^1du\phi(u,\mu)\int_0^1dx\int_0^{1-x}dy
 m_l\\\nonumber
 &&\times\left\{\left[\frac{2}{a_1}+\frac{1+(2u-1)x-y}{a_2}+\frac{1+(1-2u)x-y}{a_3}\right]
\right.\\\nonumber
&&~~~~~~\cdot \bar{u}(k_1)\gamma_5 v(k_2)
\\\nonumber  &&\left.+\int_0^{1-x-y}dz \frac{1-x-y-z}{a_4^2}
\bar{u}(k_1)\sigma^{\mu\nu}
v(k_2)\epsilon_{\mu\nu\rho\sigma}k_1^\rho k_2^\sigma \right\}.
\end{eqnarray}
The contribution of Fig.\ref{fig2}(b) can be calculated in the
same way. It can be finally shown that the contribution of
Fig.\ref{fig2}(b) $A_2$ is the same as that of Fig.\ref{fig2}(a)
if the condition $\phi(u,\mu)=\phi(\bar{u},\mu)$ maintained. This
condition is well satisfied by the wave function of $\eta_c$.

Then one can obtain the total contribution of the loop-diagrams
\begin{eqnarray}
&&A_{l}=A_1+A_2\\\nonumber &&=c_1\bar{u}(k_1)\gamma_5
v(k_2)+c_2\bar{u}(k_1)\sigma^{\mu\nu}
v(k_2)\epsilon_{\mu\nu\rho\sigma}k_1^\rho k_2^\sigma ,
\end{eqnarray}
with
\begin{eqnarray}\label{cc1}
&&c_1=\frac{Q_c^2e^4}{8\pi^2}f_{\eta_c}m_l\int_0^1du\phi(u,\mu)\int_0^1dx\int_0^{1-x}dy
 ~~~~\\\nonumber
 &&\times\left\{\left[\frac{2}{a_1}+\frac{1+(2u-1)x-y}{a_2}+\frac{1+(1-2u)x-y}{a_3}\right],
\right.\\\nonumber
\end{eqnarray}
\begin{eqnarray}\label{cc2}
&&c_2=\frac{Q_c^2e^4}{8\pi^2}f_{\eta_c}m_l\int_0^1du\phi(u,\mu)\\
&&~~~~~\cdot\int_0^1dx\int_0^{1-x}dy
 \int_0^{1-x-y}dz\frac{1-x-y-z}{a_4^2},~~~~~~\nonumber
\end{eqnarray}
where $a_1,\sim a_4$ are given in eqs. (\ref{a1to3}) and
(\ref{a4}).

The coefficients $c_1$ and $c_2$ are calculated analytically.
Before performing the integrations over Feynman parameters, let us
make the phase convention in the integrals. The imaginary angles
$\theta$'s for any imaginary quantities are restricted in the
range $-\pi<\theta <\pi$. Then the logarithms appeared during the
integration have a cut along the negative real axis. For any two
imaginary quantities $a$ and $b$, the following relations about
the logarithm are held \cite{HV}
\begin{eqnarray}
&&\ln (ab)=\ln (a)+\ln (b)+\eta(a,b),\\\nonumber && \eta(a,b)=2\pi
i\{\theta (-\mbox{Im} a)\theta (-\mbox{Im} b)\theta (\mbox{Im}
ab)\\\nonumber &&~~~~~~~~~~~~-\theta (\mbox{Im} a)\theta
(\mbox{Im} b)\theta (-\mbox{Im} ab),
\end{eqnarray}
where $\theta (x)$ is the unitstep function.

The analytical results for the Feynman parameter integration are
obtained. To express the results conveniently, some quantities are
defined as $r_c=m_c^2/m_{\eta_c}^2$, $r_l=m_l^2/m_{\eta_c}^2$,
$\alpha =\frac{2r_l}{\sqrt{1-4r_l}}$, and $\beta =\sqrt{1-4r_l}$.
The result for $c_1$ is
\begin{equation}
c_1=\frac{Q_c^2e^4}{8\pi^2}\frac{f_{\eta_c}m_l}{m_{\eta_c}^2}[f_1(u)+f_2(u)+f_3(u)],
\end{equation}
where the functions $f_1(u)$, $f_2(u)$ and $f_3(u)$ are
\begin{widetext}
\begin{eqnarray}\label{f1}
&&f_1(u)=\frac{1}{( 1 - u )u
    ( r_c + ( 1 - u )u ) }[2r_c\ln (r_c) +
    u( 1 - 2r_c - 3u + 2u^2)
     \ln (r_c - {( 1 - u ) }^2 - i\epsilon ) +
    ( 1 - u ) \\\nonumber&&~~~~\times ( u\ln (-r_c + {( 1- u ) }^2 + i\epsilon )
     - ( 2r_c + u - 2u^2 )
          \ln (r_c - u^2 - i\epsilon )+
       u\ln (-r_c + u^2 + i\epsilon ) ) ],
\end{eqnarray}
\begin{eqnarray}\label{f2}
&&f_2(u)=\left\{\frac{\beta}{(1-4r_l)^2}\right.\{2(-1+4r_l)[\ln
(2)\ln ( \frac{1 - \beta }{1 + \beta } ) -
  \ln (2)\ln ( \frac{1 + \beta }{1 - \beta } ) + 2\mbox{Li}_2(\frac{\beta }{-1 + \beta }) -
  \mbox{Li}_2(\frac{2\beta }{-1 + \beta })\\\nonumber
  &&~~~~ - 2\mbox{Li}_2(\frac{\beta }{1 + \beta }) +
  \mbox{Li}_2(\frac{2\beta }{1 + \beta })]+\frac{\beta^2}{-1+\beta^2}[\beta \ln (16)
  - 2\beta \ln (1 + \frac{1}{\beta })  + 2\beta \ln ( \frac{1}{\beta }  ) +
  2\beta \ln (\frac{1}{\beta }) \\\nonumber
  &&~~~~- 2\beta \ln (\frac{1 - \beta }{\beta }) -
  \ln (2)\ln ( \frac{1 - \beta }{1 + \beta } ) +
  \beta^2\ln (2)\ln ( \frac{1 - \beta }{1 + \beta }  )
   +
  \ln (2)\ln ( \frac{1 + \beta }{1- \beta } ) -
 {\beta }^2\ln (2)\ln ( \frac{1 + \beta }{1 - \beta }  )
   \\\nonumber
  &&~~~~ +
  2( -1 + {\beta }^2 ) \mbox{Li}_2(\frac{\beta }{-1 + \beta }) +
 ( 1- {\beta }^2 ) \mbox{Li}_2(\frac{2\beta }{-1
  + \beta }) +
  2\mbox{Li}_2(\frac{\beta }{1 + \beta })
   - 2{\beta }^2\mbox{Li}_2(\frac{\beta }{1 + \beta }) -
  \mbox{Li}_2(\frac{2\beta }{1 + \beta }) \\\nonumber
  &&~~~~
  + {\beta }^2\mbox{Li}_2(\frac{2\beta }{1 + \beta })]\}
  +\frac{-1+4r_l}{4r_l}\{ \frac{1}{1-4r_l}[- t_1\ln (-t_1)
  + t_1'\ln (-t_1') - t_2\ln (-t_2) +
  t_2'\ln (-t_2') \\\nonumber
  &&~~~~+ ( -1 + t_1 - \beta  ) \ln (1 - t_1 + \beta ) +
  ( 1 - t_1' + \beta  ) \ln (1 - t_1' + \beta ) +
  ( -1 + t_2 - \beta  ) \ln (1 - t_2 + \beta ) \\\nonumber
  &&~~~~+
  ( 1 - t_2' + \beta ) \ln (1 - t_2' + \beta )]
  + \frac{4r_l^2}{(1-4r_l)^2} [- \frac{t_1\ln (-t_1)}{\alpha ( t_1 +\alpha )}  +
  \frac{t_1'\ln (-t_1')}{\alpha ( t_1' + \alpha  ) } -
  \frac{t_2\ln (-t2)}{\alpha ( t2 + \alpha ) } +
  \frac{t_2'\ln (-t_2')}{\alpha ( t_2' + \alpha  ) }\\\nonumber
  &&~~~~ -
  \frac{\ln (\alpha )}{t_1 + \alpha } + \frac{\ln (\alpha )}{t_1' + \alpha }
  - \frac{\ln (\alpha )}{t_2 + \alpha } +
  \frac{\ln (\alpha )}{t_2' + \alpha }
  - \frac{( 1 - t_1 + \beta  ) \ln (1 - t_1 + \beta )}
   {( t_1 + \alpha  ) ( 1 + \alpha  + \beta ) } +
  \frac{( 1 - t_1' + \beta  ) \ln (1 - t_1' + \beta )}
   {( t_1' + \alpha  ) ( 1 + \alpha  + \beta  }\\\nonumber
 &&~~~~ -
  \frac{( 1 - t_2 + \beta  ) \ln (1 - t_2 + \beta )}
   {( t_2 + \alpha ) ( 1 + \alpha  + \beta ) } +
  \frac{( 1 - t_2' + \beta  ) \ln (1 - t_2' + \beta )}
   {( t_2' + \alpha  )( 1 + \alpha  + \beta ) }
   + \frac{\ln (1 + \alpha  + \beta )}{t_1 + \alpha } -
  \frac{\ln (1 + \alpha  + \beta )}{t_1' + \alpha }\\\nonumber
&&~~~~
   + \frac{\ln (1 + \alpha  + \beta )}{t_2 + \alpha } -
  \frac{\ln (1 + \alpha  + \beta )}{t_2' + \alpha }]
  + \frac{1}{{\beta }^3( 2r_l + \beta  + {\beta }^2 ) }
  \ln (\frac{2r_l +( -1 + \sqrt{1 - 4r_l} ) u}
     {-2r_l + ( 1 + \sqrt{1 - 4r_l} ) u})[\beta ( 1 + \beta) ( 4r_l\\\nonumber
  &&~~~~ + \beta
      + {\beta }^2)  +
  4r_l( 2r_l + \beta  + {\beta }^2 ) \ln (2r_l) -
  4r_l( 2r_l + \beta  + {\beta }^2 ) \ln (2r_l + \beta  + {\beta }^2)] \\\nonumber
  &&~~~~-\frac{4r_l}{(1-4r_l)^{3/2}}[\ln (-t_1)\ln (\frac{\alpha }{t_1 + \alpha })
  - \ln (-t_1')\ln (\frac{\alpha }{t_1' + \alpha }) +
  \ln (-t_2)\ln (\frac{\alpha }{t_2 + \alpha })\\\nonumber
  &&~~~~
  - \ln (-t_2')\ln (\frac{\alpha }{t_2'+ \alpha }) -
  \ln (1 - t_1 + \beta )\ln (\frac{1 + \alpha  + \beta }{t_1 + \alpha }) +
  \ln (1 - t_1' + \beta )\ln (\frac{1 + \alpha  + \beta }{t_1'+ \alpha }) \\\nonumber
&&~~~~-
  \ln (1 - t_2 + \beta )\ln (\frac{1 + \alpha  + \beta }{t_2 + \alpha }) +
  \ln (1 - t_2p + \beta )\ln (\frac{1 + \alpha  + \beta }{t_2' + \alpha }) +
  \mbox{Li}_2(\frac{t_1}{t_1 + \alpha }) - \mbox{Li}_2(\frac{t_1'}{t_1' + \alpha }) \\\nonumber
  &&~~~~+
  \mbox{Li}_2(\frac{t_2}{t_2 + \alpha }) - \mbox{Li}_2(\frac{t_2'}{t_2' + \alpha }) -
  \mbox{Li}_2(\frac{-1 + t_1 - \beta }{t_1 + \alpha }) +
  \mbox{Li}_2(\frac{-1 + t_1' - \beta }{t_1' + \alpha }) -
  \mbox{Li}_2(\frac{-1 + t_2 - \beta }{t_2 + \alpha }) \\\nonumber
  &&~~~~+
  \mbox{Li}_2(\frac{-1 + t_2' - \beta }{t_2' + \alpha })]  \}
  +\frac{i\pi}{2r_l}+\frac{1}{2r_l(r_l+(-1+u)u)}[r_l\ln(-r_l)\\\nonumber
  &&~~~~
  +(r_l+(-1+u)u)\ln((1-u)u)-r_l\ln((-1+u)u)-(r_l-u+u^2)\ln(u^2)]\left.\frac{}{}\right\},
\end{eqnarray}
\end{widetext}
where
\begin{eqnarray}
&&t_1=\frac{2r_l( {\sqrt{1 - 4r_l}} + u ) +2{r_l(1-u)}}{2r_l - ( 1
+ {\sqrt{1 - 4r_l}} ) u} +
  i\epsilon ,\\
&&t_2=\frac{ 2r_l( {\sqrt{1 - 4r_l}} + u )-2{r_l(1-u)}}{2r_l - ( 1
+ {\sqrt{1 - 4r_l}} ) u} -
  i\epsilon ,\\
&&t_1'=\frac{2r_l( {\sqrt{1 - 4r_l}} - u ) -2{r_l(1-u)}}{2r_l +(
-1 + {\sqrt{1 - 4r_l}} ) u} +
  i\epsilon ,\\
&&t_2'=\frac{2r_l( {\sqrt{1 - 4r_l}} - u ) +2{r_l(1-u)}}{2r_l +(
-1 + {\sqrt{1 - 4r_l}} ) u} -
  i\epsilon .
\end{eqnarray}
The function $\mbox{Li}_2(z)$ is the polylogarithm function, which
is defined as $\mbox{Li}_2(z)=\int_z^0 dt\frac{\ln (1-t)}{t}$.
Finally, the third function is
\begin{equation}\label{f3}
f_3(u)=f_2(1-u).
\end{equation}

The analytical result of the 3-fold Feynman parameter integration
in $c_2$ (eq.(\ref{cc2})) is tedious, it is not presented here.
Some formulas used in the procedure of integration are given in
Appendix B.

It is not difficult to calculate the contribution of weak
interaction, where the virtual $Z^0$ acts as the intermediate
state (Fig.\ref{fig2}(c)). The contribution of Fig.\ref{fig2}(c)
to the amplitude of $\eta_c\to l^+l^-$ is
\begin{eqnarray}
A_{z^0}=\langle l^+l^-|\bar{l} V_{\bar{l}l
Z}l\frac{-i}{p^2-m_Z^2+i\epsilon} \bar{c} V_{\bar{c}c
Z}c|\eta_c\rangle ,
\end{eqnarray}
where $p=k_1+k_2$, and $V_{\bar{l}l Z}$ and $V_{\bar{c}c Z}$ are
the coupling vertices of $\bar{l}l Z$ and $\bar{c}c Z$,
respectively,
\begin{eqnarray*}
&&V_{\bar{l}l Z}=ie\gamma^\mu
\left(\frac{-4\sin^2\theta_W+1}{4\sin\theta_W\cos\theta_W}
-\frac{1}{4\sin\theta_W\cos\theta_W}\gamma_5\right),\\
&&V_{\bar{c}c Z}=ie\gamma^\mu
\left(\frac{8\sin^2\theta_W-3}{12\sin\theta_W\cos\theta_W}
+\frac{1}{4\sin\theta_W\cos\theta_W}\gamma_5\right),\\
\end{eqnarray*}
with $\theta_W$ being the Weinberg angle.

Contracting the leptonic final state with the leptonic fields
operator and using the definition of the decay constant of
$\eta_c$, one can finally obtain the amplitude
\begin{equation}
A_{z^0}=\frac{2m_le^2}{(4\sin\theta_W\cos\theta_W)^2}\frac{f_{\eta_c}}{p^2-m_Z^2}
 \bar{u}(k_1)\gamma_5v(k_2) .
\end{equation}
Then the amplitude of the leptonic decay $\eta_c\to\l^+\l^-$ is
the sum of the loop and tree diagrams
\begin{equation}
A(\eta_c\to\l^+\l^-)=A_l+A_{Z^0}.
\end{equation}
The decay width of this decay is
\begin{equation}
\Gamma(\eta_c\to\l^+\l^-)=\frac{1}{8\pi}|A(\eta_c\to\l^+\l^-)|^2
\frac{|\vec{k}|}{m_{\eta_c}^2},
\end{equation}
where $\vec{k}$ is the three-momentum of one of the leptons in the
rest frame of $\eta_c$. The branching ratio of the decay is
defined by
\begin{equation}
 Br(\eta_c\to\l^+\l^-)=
 \frac{\Gamma(\eta_c\to\l^+\l^-)}{\Gamma_{\mbox{tot}}},
\end{equation}
where $\Gamma_{\mbox{tot}}$ are the total decay width of the
$\eta_c$ meson.

In the numerical calculation, the wave function of $\eta_c$ is
taken to be \cite{wv}
\begin{equation}
\phi(u,\mu\sim
m_c)=N\;4u(1-u)\mbox{exp}\left(-\frac{\beta}{4u(1-u)}\right),
\end{equation}
where $N$ is the normalization factor, the parameter $\beta
=3.8\pm 0.7$. The value of the decay constant of $\eta_c$ is
$f_{\eta_c}=0.346\pm 0.017~\mbox{GeV}$ \cite{wv}, the mass of $c$
quark $m_c=1.4~\mbox{GeV}$, the total decay width of $\eta_c$ is
$\Gamma_{\mbox{tot}}=26.7\pm 3.0 ~\mbox{MeV}$ \cite{PDG2008}.

With the parameter inputs and the wave function of $\eta_c$ given
above, the prediction to the branching ratio of
$\eta_c\to\gamma\gamma$ is
\begin{equation}\label{b1}
B(\eta_c\to\gamma\gamma)=2.43^{+0.39}_{-0.34}\times 10^{-4},
\end{equation}
where the uncertainty comes from the uncertainty of the total
decay width of $\eta_c$, the uncertainty of the decay constant
$f_{\eta_c}$ and the uncertainty of the parameter $\beta$ in the
wave function. The main contribution comes from the uncertainties
of the total decay width and decay constant of $\eta_c$, which are
about 10\% of the central value, while the error caused by the
parameter $\beta$ is small, which is only about 1\%.

The prediction in eq.(\ref{b1}) is in good agreement with the
experimental data
\begin{equation}\label{b1exp}
B^{\mbox{exp}}(\eta_c\to\gamma\gamma)=2.4^{+1.1}_{-0.9}\times
10^{-4}.
\end{equation}

Next I will go on to the numerical discussion on $\eta_c\to
l^+l^-$.

\begin{figure}[hbt]
 \epsfig{file=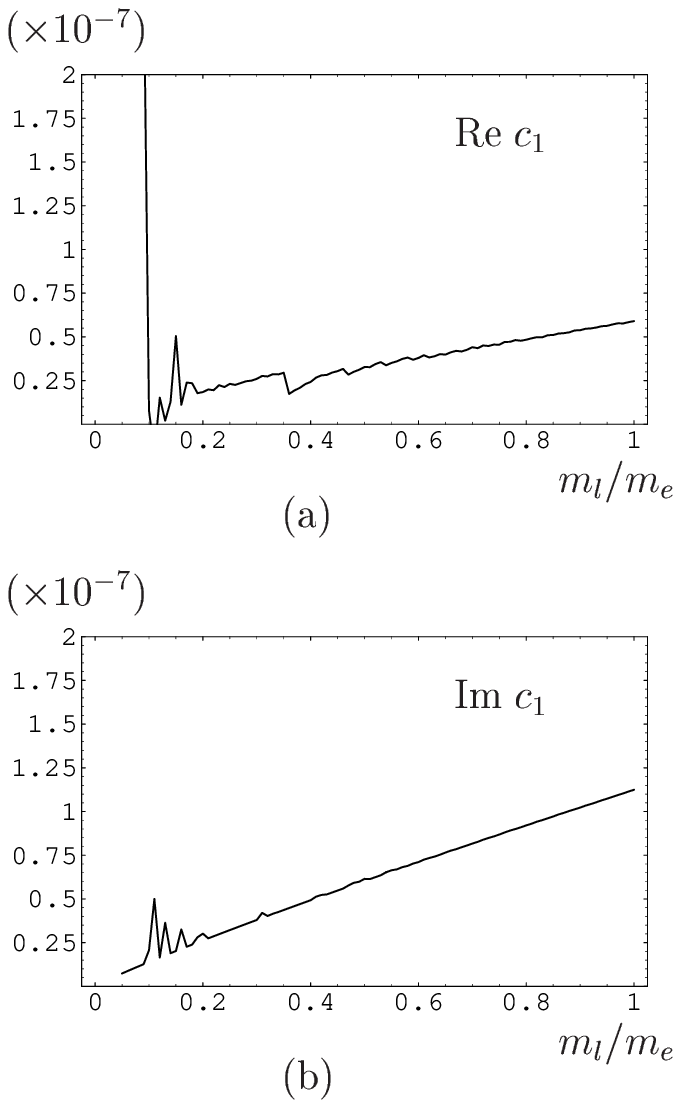, width=6cm,height=9cm}
\caption{\small Behavior of the coefficient $c_1$ as $m_l\to 0$.
(a) the real part of $c_1$; (b) the imaginary part of $c_1$. The
horizontal axis is for the ratio of $m_l/m_e$. } \label{fig3}
\end{figure}

The loop integral is infrared divergent if the lepton mass is
zero. This can be shown by taking the asymptotic limit of $r_l\to
0$ for the analytical results of loop-integral-functions $f_1(u)$,
$f_2(u)$ and $f_3(u)$ in eqs. (\ref{f1}), (\ref{f2}) and
(\ref{f3}). The most singular term behaves like $\frac{1}{r_l}\ln
(r_l)$. Considering the decay amplitude in eq. (\ref{ax1}) is
proportional to $m_l\sim \sqrt{r_l}$ due to the helicity
suppression, the most singular term in the decay amplitude behaves
like $\frac{1}{\sqrt{r_l}}\ln (r_l)$. The curves for the
coefficient $c_1$ as the lepton mass approaching zero are shown in
Fig.\ref{fig3}. Only when the lepton mass $m_l<0.2m_e$, here $m_e$
is the mass of electron, does the numerical result begin to be
severely affected by the infrared divergence. As the lepton mass
not less than $0.2m_e$, the numerical value is reliable. The
situation for $c_2 $ is similar to $c_1$. Therefore, for the cases
of $m_l=m_e$ and $m_l=m_\mu$ ($m_\mu$ the muon mass), the loop
integral is not affected by the infrared divergence.

The values of $c_1$ and $c_2$ are presented in Table \ref{tab1}.

\begin{widetext}
\begin{center}
\begin{table}[htbp]
  \caption{The coefficient $c_1$ and $c_2$ in the decay amplitudes
  of $\eta_c\to e^+e^-$ and $\eta_c\to \mu^+\mu^-$.}
   \begin{tabular}{c|c|c} \hline\hline
    &  $c_1$  & $c_2$ \\ \hline
 $\eta_c\to e^+e^-$ & $5.90\times 10^{-8}+1.12\times 10^{-7}i$&
    $-2.75\times 10^{-8}-1.43\times 10^{-9}i$  \\ \hline
 $\eta_c\to \mu^+\mu^-$ & $3.20\times 10^{-6}+1.02\times 10^{-5}i
    $ &$-5.53\times 10^{-7}+5.36\times 10^{-6}i$  \\
   \hline \hline
 \end{tabular}
  \label{tab1}
\end{table}
  \end{center}
\end{widetext}

The branching ratios are given in Table \ref{tab2}. The leptonic
decays of $\eta_c$ are dominated by the two-photon loop diagrams,
the contribution of $Z^0$ tree diagram modifies the decay rates by
less than 1\%.

\begin{center}
\begin{table}[htbp]
  \caption{The branching ratios of $\eta_c\to e^+e^-$
  and $\eta_c\to \mu^+\mu^-$. The column ``Only loop" means
  the contribution of only loop diagram, without the contribution
  of $Z^0$ tree diagram, while ``Loop+$Z^0$" denotes the contribution
  of both loop and $Z^0$ tree diagrams.}
   \begin{tabular}{c|c|c} \hline\hline
    & Only loop  & Loop+$Z^0$ \\ \hline
 $\eta_c\to e^+e^-$ & $4.77^{+0.77}_{-0.66}\times 10^{-13}$&
    $4.74^{+0.77}_{-0.66}\times 10^{-13}$  \\ \hline
 $\eta_c\to \mu^+\mu^-$ & $6.41^{+1.04}_{-0.89}\times 10^{-9}$
  & $6.39^{+1.03}_{-0.89}\times 10^{-9}$  \\
   \hline \hline
 \end{tabular}
  \label{tab2}
\end{table}
  \end{center}

\begin{figure}[hbt]
\begin{center}
\vspace{1CM}
 \epsfig{file=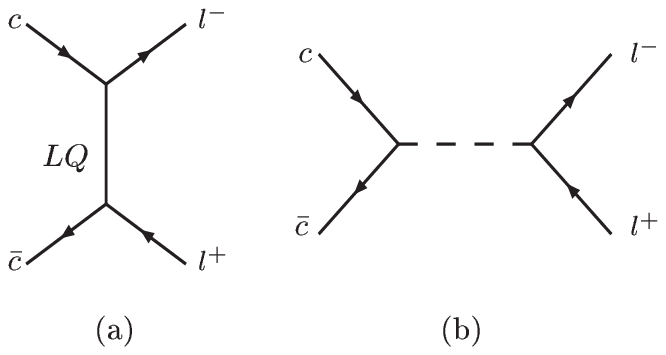, width=6.cm,height=3.5cm}
 \end{center}
\vspace{-0.5cm} \caption{\small Diagrams for $\eta_c\to l^+l^-$ in
physics beyond the standard model. (a) leptoquark contribution,
and (b) light pseudoscalar Higgs contribution.} \label{fig4}
\end{figure}

The branching ratios of $\eta_c\to l^+l^-$ are very tiny within
the frame work of the standard model. Charm physics including
leptonic, semileptonic and hadronic charm decays will be studied
at BESIII \cite{hb,bes3}. About $10\times 10^9$ $J/\psi$ events
will be accumulated with one year designed luminosity. Considering
the branching ratio of $J/\psi\to\gamma\eta_c$ being $(1.3\pm
0.4)\%$ \cite{PDG2008}, about $1.3\times 10^8$ $\eta_c$ mesons can
be produced through the radiative decays of
$J/\psi\to\gamma\eta_c$. Assuming the detection efficiency is
about 20\%, then the sensitivity of the measurement of $\eta_c\to
l^+\l^-$ can reach $3.8\times 10^{-8}$. The standard model
prediction to the leptonic decay of $\eta_c\to l^+\l^-$ is just
below the sensitivity of BESIII. However, the low decay rate
within the standard model makes the leptonic decay of $\eta_c\to
l^+\l^-$ sensitive to physics beyond the standard model, such as
the existence of leptoquark bosons, Fig. \ref{fig4}(a), or the
light pseudoscalar Higgs boson, Fig. \ref{fig4}(b). Especially
when the mass of the light pseudoscalar Higgs lies near the mass
of $\eta_c$ meson, Fig. \ref{fig4}(b) can significantly enhance
the leptonic decay rate of $\eta_c$. Measurement of $\eta_c\to
l^+l^-$ decay, especially for $\eta_c\to \mu^+\mu^-$ channel, can
give information of physics beyond the standard model.

In summary, the radiative and leptonic decays of
$\eta_c\to\gamma\gamma$ and $\eta_c\to l^+l^-$ have been studied
consistently within the framework of the standard model. The
theoretical prediction of the decay rate of
$\eta_c\to\gamma\gamma$ is in well agreement with the experimental
measurement. For the leptonic decays of $\eta_c\to \mu^+\mu^-$ and
$e^+e^-$, within the framework of the standard model, the loop
diagrams of electromagnetic interaction dominate, the tree diagram
of weak interaction involving $z^0$ propagator can only modifies
the decay rates by less than 1\%. The decay rates of $\eta_c\to
\mu^+\mu^-$ and $e^+e^-$ are tiny, which makes them sensitive to
some new physics beyond the standard model. Measurement of these
decay rates may shed light on the existence of new physics.

I thank H.B. Li for helpful discussions, and be grateful to Kavli
Institute for Theoretical Physics China, where part of this work
was done. This work is supported in part by the National Natural
Science Foundation of China under contracts Nos. 10575108,
10735080.

\begin{center} Appendix A \end{center}
The wave function of a meson $M$ composed of a quark-antiquark
pair $\bar{q}_1q_2$ can be defined through the matrix element
$\langle 0| \bar{q}_1(x)_\alpha q_2(y)_\beta |M\rangle$, where
summation over color degrees of freedom and a gauge invariant
factor $P\mbox{exp}\{i\int dz^\mu T^aA^a_\mu (z)\}$ between
$\bar{q}_1(x)$ and $q_2(y)$ are indicated.

Using the equality of Fietz transformation
\begin{eqnarray}
&&\delta_{ik}\delta_{lj}=\frac{1}{4}\delta_{ij}\delta_{lk}
+\frac{1}{4}(\gamma_5)_{ij}(\gamma_5)_{lk}\\\nonumber &&
+\frac{1}{4}(\gamma_\mu)_{ij}(\gamma^\mu)_{lk}
-\frac{1}{4}(\gamma_\mu\gamma_5)_{ij}(\gamma^\mu\gamma_5)_{lk}\\\nonumber
&&
+\frac{1}{8}(\sigma_{\mu\nu}\gamma_5)_{ij}(\sigma^{\mu\nu}\gamma_5)_{lk},
\end{eqnarray}
the matrix element can be written as
\begin{eqnarray}
&&\langle 0| \bar{q}_1(x)_\alpha q_2(y)_\beta |M\rangle
\\\nonumber &=&\langle 0| \bar{q}_1(x)_\rho q_2(y)_\sigma |M\rangle
\delta_{\rho\alpha}\delta_{\beta\sigma}\\\nonumber &=&\langle 0|
\bar{q}_1(x)_\rho q_2(y)_\sigma |M\rangle
(\frac{1}{4}\delta_{\rho\sigma}\delta_{\beta\alpha}
+\frac{1}{4}(\gamma_5)_{\rho\sigma}(\gamma_5)_{\beta\alpha}\\\nonumber
&& +\frac{1}{4}(\gamma_\mu)_{\rho\sigma}(\gamma^\mu)_{\beta\alpha}
-\frac{1}{4}(\gamma_\mu\gamma_5)_{\rho\sigma}(\gamma^\mu\gamma_5)_{\beta\alpha}\\\nonumber
&&
+\frac{1}{8}(\sigma_{\mu\nu}\gamma_5)_{\rho\sigma}(\sigma^{\mu\nu}\gamma_5)_{\beta\alpha})
\\\nonumber
&=&\frac{1}{4}\langle 0| \bar{q}_1(x) q_2(y)|M\rangle
\delta_{\beta\alpha}+\frac{1}{4}\langle 0| \bar{q}_1(x)\gamma_5
q_2(y)|M\rangle (\gamma_5)_{\beta\alpha}\\\nonumber &&
+\frac{1}{4}\langle 0| \bar{q}_1(x)\gamma_\mu q_2(y)|M\rangle
(\gamma^\mu)_{\beta\alpha}\\\nonumber &&-\frac{1}{4}\langle 0|
\bar{q}_1(x)\gamma_\mu \gamma_5 q_2(y)|M\rangle
(\gamma^\mu\gamma_5)_{\beta\alpha}\\\nonumber &&
+\frac{1}{8}\langle 0| \bar{q}_1(x)\sigma_{\mu\nu} \gamma_5
q_2(y)|M\rangle (\sigma^{\mu\nu}\gamma_5)_{\beta\alpha}.
\end{eqnarray}
For a pseudoscalar meson $M$, the matrix elements of the scalar
and vector current are all zero. For the nonzero matrix element
the wave function can be defined as \cite{bf}
\begin{equation}
\langle 0| \bar{q}_1(x)\gamma_\mu \gamma_5 q_2(y)|M\rangle
 =if_Mp_\mu\int_0^1 du e^{-i(up\cdot x+\bar{u}p\cdot
y)}\phi(u,\mu),
\end{equation}
\begin{equation}
\langle 0| \bar{q}_1(x)i \gamma_5 q_2(y)|M\rangle
=\frac{f_Mm^2_M}{m_1+m_2}\int_0^1 du e^{-i(up\cdot x+\bar{u}p\cdot
y)}\phi_p(u,\mu),
\end{equation}
\begin{eqnarray}
\langle 0| \bar{q}_1(x)\sigma_{\mu\nu} \gamma_5 q_2(y)|M\rangle
&=&i\frac{f_M m^2_M}{m_1+m_2}(p_\mu z_\nu-p_\nu z_\mu) \\\nonumber
&&\cdot\int_0^1 du e^{-i(up\cdot x+\bar{u}p\cdot
y)}\frac{\phi_\sigma(u,\mu)}{6},
\end{eqnarray}
where $z=y-x$, $\mu$ is the energy scalar where the perturbative
and non-perturbative dynamics of QCD can be factorized, $f_M$ the
decay constant, $p_\mu$ the momentum of the meson $M$, and
$m_{M}$, $m_1$ and $m_2$ are the masses of the meson $M$, quarks
$q_1$ and $q_2$, respectively. The wave function $\phi(u,\mu)$ is
of twist-2, and $\phi_p(u,\mu)$ and $\phi_\sigma(u,\mu)$ are of
twist-3.

Then the matrix element $\langle 0| \bar{q}_1(x)_\alpha
q_2(y)_\beta |M\rangle$ becomes
\begin{eqnarray}
&&\langle 0| \bar{q}_1(x)_\alpha q_2(y)_\beta |M\rangle\\\nonumber
&&=-\frac{i}{4}f_{M}\int_0^1 du e^{-i(up\cdot x+\bar{u}p\cdot
y)}[\not {p}\gamma_5\phi(u,\mu)\\\nonumber
&&+\frac{m^2_M}{m_1+m_2}\gamma_5\phi_p(u,\mu)
-\frac{1}{2}\frac{m^2_M}{m_1+m_2}(p_\mu z_\nu\\\nonumber &&-p_\nu
z_\mu)\sigma_{\mu\nu}\frac{\phi_\sigma(u,\mu)}{6}
\gamma_5]_{\beta\alpha}.
\end{eqnarray}
For the decay processes of $\eta_c$ considered in this work, only
the leading-twist wave function is considered. The contributions
of the wave functions of twist-3 are suppressed due to their spin
structure. Therefore they are safely dropped.

\begin{center} Appendix B \end{center}
Some formulas used in the calculation of the 3-fold Feynman
parameter integration are
\begin{eqnarray}
&&\int dx\frac{1}{ax^2+bx+c}=\frac{1}{\sqrt{b^2-4ac}}[~\ln
(2ax+b~~~~~~~\\\nonumber&&~~~~~-\sqrt{b^2-4ac}~]-\ln
(2ax+b+\sqrt{b^2-4ac}~ ],
\end{eqnarray}
\begin{eqnarray}
&&\int
dx\frac{x}{(ax^2+bx+c)^2}=\frac{bx+2c}{(b^2-4ac)(ax^2+bx+c)}~~~~~~~~~~~
\\\nonumber&&~~~~~~~~~+\frac{b}{b^2-4ac}\int
dx\frac{1}{ax^2+bx+c}.
\end{eqnarray}
For the integration of the type
\begin{equation}\label{inty4}
\int dy \frac{n_1y+n_2}{(a_1y^2+b_1y+c_1)(a_2y^2+b_2y+c_2)},
\end{equation}
the integrand can be decomposed as
\begin{eqnarray}
&&\frac{n_1y+n_2}{(a_1y^2+b_1y+c_1)(a_2y^2+b_2y+c_2)}\\\nonumber
&&=\frac{k_1y+l_1}{a_1y^2+b_1y+c_1}+\frac{k_2y+l_2}{a_2y^2+b_2y+c_2},
\end{eqnarray}
where the coefficients $k_{1,2}$ and $l_{1,2}$ can be solved
within the above equation. Then the integration of eq.
(\ref{inty4}) can be reduced into a simpler form.

For the integral like
\begin{equation}
\int dx \frac{k x+l}{ax^2+bx+c}\frac{1}{\sqrt{e x+f}}\ln(g x+b\pm
\sqrt{e x+f}),
\end{equation}
the integration can be performed by making the variable
transformation $t=\sqrt{ex+f}$. While for
\begin{eqnarray}
&&\int dx \frac{k
x+l}{a_1x^2+b_1x+c_1}\frac{1}{\sqrt{a_2x^2+b_2x+c_2}}\\\nonumber&&
\times\ln(gx+b\pm \sqrt{a_2x^2+b_2x+c_2}),
\end{eqnarray}
with $a_2>0$, the variable transformation shall be
$\sqrt{a_2x^2+b_2x+c_2}=t-\sqrt{a_2} x$.

\begin{eqnarray}
\int
dt\frac{\ln(t-a)}{t-b}&=&\ln(t-a)\ln\left(\frac{t-a}{a-b}+1\right)
\\\nonumber &&+\mbox{Li}_2\left(-\frac{t-a}{a-b}\right),
\end{eqnarray}
where $\mbox{Li}_2(z)$ is the polylogarithm function, which is
defined as $\mbox{Li}_2(z)=\int_z^0\frac{\ln (1-t)}{t}$.



\end{document}